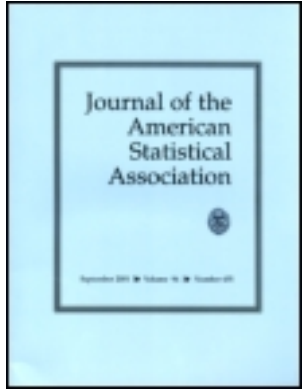

# Building Consistent Regression Trees From Complex Sample Data

Daniell Toth and John L. Eltinge
Daniell Toth is Mathematical Statistician, Office of Survey Methods Research, Bureau of Labor Statistics, Suite 1950, Washington, DC 20212. John L. Eltinge is Associate Commissioner, Office of Survey Methods Research, Bureau of Labor Statistics, Suite 1950, Washington, DC 20212. The views expressed on statistical, methodological, technical, and operational issues are those of the authors and not necessarily those of the U.S. Bureau of Labor Statistics. The authors thank the editors and referees for their very helpful comments that substantially improved this article. We also thank our colleague Michail Sverchkov for his helpful comments.

PLEASE SCROLL DOWN FOR ARTICLE

# Building Consistent Regression Trees From Complex Sample Data

Daniell TOTH and John L. ELTINGE

In the past several years a wide range of methods for the construction of regression trees and other estimators based on the recursive partitioning of samples have appeared in the statistics literature. Many applications involve data collected through a complex sample design. At present, however, relatively little is known regarding the properties of these methods under complex designs. This article proposes a method for incorporating information about the complex sample design when building a regression tree using a recursive partitioning algorithm. Sufficient conditions are established for asymptotic design $L^2$ consistency of these regression trees as estimators for an arbitrary regression function. The proposed method is illustrated with Occupational Employment Statistics establishment survey data linked to Quarterly Census of Employment and Wage payroll data of the Bureau of Labor Statistics. Performance of the nonparametric estimator is investigated through a simulation study based on this example.

KEY WORDS: Analysis of complex survey data; Asymptotic consistency; Classification and Regression Trees (CART); Complex sample design; Nonparametric regression; Recursive partitioning.

## 1. INTRODUCTION

A central problem in analyzing survey data is using available auxiliary information to estimate finite population parameters of interest. The functional relationship between the auxiliary variables and a finite population parameter is often quite complex, involving a number of interaction effects. One generally prefers to use nonparametric models which provide good approximations for a larger class of functions than a given parametric model and are usually more robust under general sampling designs. Work with such models is now an emerging area of importance for analyzing survey data (Dorfman and Hall 1993; Breidt and Opsomer 2000, 2009; Zheng and Little 2004; Breidt, Claeskens, and Opsomer 2005).

One important type of nonparametric model is the binary tree built from a recursive partitioning algorithm applied to the dataset. Since Morgan and Sonquist (1963) suggested using these simple nonparametric models for analyzing survey data, they have grown in importance (Göksel, Judkins, and Mosher 1992; De'ath and Fabricius 2000; De'ath 2002; Lobell et al. 2005; Benedetti, Espa, and Lafratta 2008; Mendez et al. 2008). The use of this method in survey analysis has largely ignored design information such as the selection probabilities of individual observations and has been done in the absence of any theoretical justification (Simpson et al. 2003); this can lead to biased and misleading results (Holt, Smith, and Winter 1980; Nathan and Holt 1980). In addition, Edwards et al. (2006) considered the impact of purposive designs on regression tree analysis.

The aim of this article is to provide some theoretical justification for use of these procedures on survey data. We provide a set of sufficient conditions on the population, the sample design, and on the partitioning algorithm for the proposed estimator to be asymptotically design unbiased (ADU) and asymptotically design consistent (ADC) with respect to the superpopulation model (Isaki and Fuller 1982; Robinson and Särndal 1983). More specifically, for each value $x$ we define an estimator $\tilde{h}(\mathbf{x})$ based on a class of recursive partitioning algorithms that is ADU and ADC for the superpopulation value $h(x)$.

One important complication in working with recursive partitioning algorithms is that the estimator is based on a sample driven partition. Given a partition, the proofs that follow are in line with the standard estimating equation approach. That is, the function is written as a function of population totals and then a design consistent estimator for the totals is used to derive an estimator for the function. However, the fact that the partition is dependent on the sample drawn requires that the arguments account for this source of randomness.

This is done by extending the work in a series of articles by Stone (1977) and Gordon and Olshen (1978, 1980). Those articles present conditions for a recursive partitioning algorithm to produce an $L^p$ consistent estimator of a regression model when the observations are independent and identically distributed (iid) from a general distribution. This article provides conditions on the sampling design and the recursive partitioning algorithm such that: (1) the recursive partition applied to the sample data satisfies the conditions of Gordon and Olshen (1978, 1980) for the population estimator based on this partition to be $L^p$ consistent with respect to the superpopulation model; and (2) the estimator based on the sample data is $L^2$ consistent with respect to the design as an estimator for the finite population quantity defined using the same partition. Therefore, after arguing that the finite population quantity defined for the sample based partition satisfies the conditions for asymptotic consistency, the arguments follow those of the estimating equation approach.

The next section reviews some of the primary results of Gordon and Olshen (1978, 1980) restated in terms of an iid finite population $U_\nu$ generated from a superpopulation model $\xi$. Section 3 establishes the asymptotic design unbiasedness and consistency of the sample based estimator. Section 4 provides

Daniell Toth is Mathematical Statistician, Office of Survey Methods Research, Bureau of Labor Statistics, Suite 1950, Washington, DC 20212 (E-mail: *toth.daniell@bls.gov*). John L. Eltinge is Associate Commissioner, Office of Survey Methods Research, Bureau of Labor Statistics, Suite 1950, Washington, DC 20212 (E-mail: *eltinge.john@bls.gov*). The views expressed on statistical, methodological, technical, and operational issues are those of the authors and not necessarily those of the U.S. Bureau of Labor Statistics. The authors thank the editors and referees for their very helpful comments that substantially improved this article. We also thank our colleague Michail Sverchkov for his helpful comments.

In the Public Domain
**Journal of the American Statistical Association**
**December 2011, Vol. 106, No. 496, Theory and Methods**
DOI: 10.1198/jasa.2011.tm10383

an example using a recursive partitioning algorithm to analyze Occupational Employment Statistics (OES) Survey establishment data linked to Quarterly Census of Employment and Wage (QCEW) payroll data of the Bureau of Labor Statistics (BLS). This example demonstrates the effect of complex sample designs on regression tree models. Simulation results for the method applied to repeated samples of these data using the original probability of selection and using probability proportional to size (pps) are also given in this section. Section 5 summarizes the results of the article, discusses other findings, and considers several possible directions for future research. Appendix A contains some proofs.

## 2. RECURSIVE PARTITIONING

Consider a population $\{(Y_1, \mathbf{X}_1), \ldots, (Y_{N_\nu}, \mathbf{X}_{N_\nu})\}$, indexed by the set $U_\nu = \{1, \ldots, N_\nu\}$. The elements of the set are independently and identically distributed according to the model $\xi$ with distribution function $F(y, \mathbf{x})$. Here $y$ is univariate and $\mathbf{x}$ is a $d$-dimensional vector. We use $F_Y(y)$ and $F_l(x_l)$ to denote the univariate marginal distributions of $Y$ and $X_l$, respectively. The notation $F^-(x)$ is used to represent $\lim_{\epsilon \to 0^-} F(x + \epsilon)$ for any univariate cdf $F(\cdot)$. Gordon and Olshen (1978, 1980) noted that the arguments that follow can be proved without loss of generality for $\mathbf{X} \in (0, 1)^d$. The same is true for the algorithms and proofs that follow, so we also assume $\mathbf{X}$ is of this form.

Let $E_\xi$ denote the expectation evaluated with respect to $\xi$. Now, consider estimation of a superpopulation regression function $h(\mathbf{x}) = E_\xi[Y|\mathbf{X} = \mathbf{x}]$. For a set $A$ that intersects the support of the auxiliary variable $\mathbf{X}$, define the function $h(A) = E_\xi[Y|\mathbf{X} \in A]$. Gordon and Olshen (1978, 1980) considered an estimator $\hat{h}_{N_\nu}(\mathbf{x})$ constructed from a recursive partitioning algorithm. They gave sufficient conditions for these algorithms to satisfy

$$\lim_{N_\nu \to \infty} E_\xi\left[\left|h(\mathbf{X}) - \hat{h}_{N_\nu}(\mathbf{X})\right|^p\right] = 0 \quad (1)$$

for a fixed $p > 1$. The remainder of this section reviews results from those articles.

A recursive partitioning algorithm begins by splitting the entire sample into two subsets according to the value of one of the auxiliary variables $X_l$ for $l = 1, \ldots, d$. For example, the sample could be divided into the two sets $\{i \in U_\nu | x_{il} \leq c\}$ and $\{i \in U_\nu | x_{il} > c\}$ for some constant $c$. The mean estimator is calculated on each subset separately. This procedure is repeated on each subset until the entire dataset is partitioned into boxes such that each box contains less than a predefined number of elements. At each step, the split that results in the largest decrease in the estimated mean squared error for the estimator applied to the given dataset is chosen from among all possible splits on the auxiliary variables.

Let $Q^{N_\nu}$ be the set of partitioning boxes resulting from the algorithm. Let supp($\mathbf{X}$) be the support of the random vector $\mathbf{X}$. Then $B^{N_\nu}$ denotes an arbitrary box in $Q^{N_\nu}$ and $B^{N_\nu}(\mathbf{x})$ represents the box in the partition containing the value $\mathbf{x} \in$ supp($\mathbf{X}$). A box $B^{N_\nu}$ in a given partition $Q^{N_\nu}$ has two corresponding index vectors $\mathbf{a}(B^{N_\nu}) = (a_1(B^{N_\nu}), \ldots, a_d(B^{N_\nu}))$ and $\mathbf{b}(B^{N_\nu}) = (b_1(B^{N_\nu}), \ldots, b_d(B^{N_\nu}))$. These vectors define the contents of the box to the extent that every $\mathbf{x} \in B^{N_\nu}$ satisfies $a_l(B^{N_\nu}) \leq x_l \leq b_l(B^{N_\nu})$ for $l = 1, \ldots, d$.

Next we describe a norm on a given partition of the auxiliary variable space. If $F$ is the given distribution of $\mathbf{X}$, let $P$ denote the corresponding probability measure and $F_l$ the corresponding marginal distribution of component $x_l$. Define the *l-norm of partition $Q^{N_\nu}$ relative to $F$*

$$\|Q^{N_\nu}\|_l^F = \sum_{B^{N_\nu} \in Q^{N_\nu}} \left\{\left[F_l(b_l(B^{N_\nu})) - F_l(a_l(B^{N_\nu}))\right]P(B^{N_\nu})\right\} \quad (2)$$

and the *l-norm of partition $Q^{N_\nu}$ relative to $F^-$*

$$\|Q^{N_\nu}\|_l^{F^-} = \sum_{B^{(N_\nu)} \in Q^{N_\nu}} \left\{\left[F_l^-(b_l(B^{N_\nu})) - F_l^-(a_l(B^{N_\nu}))\right]P(B^{N_\nu})\right\}. \quad (3)$$

These measures are roughly the expected probability mass assigned to a box in $Q^{N_\nu}$ with respect to the distributions $F_l$ or $F_l^-$. For example, a partition $Q^{N_\nu}$ that includes many splits on $X_l$ will tend to have a smaller $\|Q^{N_\nu}\|_l^F$ and $\|Q^{N_\nu}\|_l^{F^-}$ than a partition with fewer splits on $X_l$. See Appendix B for some related comments on the $\|Q^{N_\nu}\|_l^F$ and $\|Q^{N_\nu}\|_l^{F^-}$ measures.

Next we define some final notation used in the statement of the main result of Gordon and Olshen (1980) that we will be extending. For a given value of $Y = y$ and a set $A \subset$ supp($\mathbf{X}$) define the empirical conditional distribution function

$$\hat{F}_{N_\nu}(y|A) = \left(\sum_{i \in U_\nu} \mathbb{1}_{\{\mathbf{x}_i \in A\}}\right)^{-1} \sum_{i \in U_\nu} \mathbb{1}_{\{y_i < y\}} \mathbb{1}_{\{\mathbf{x}_i \in A\}}$$

and the empirical probability

$$\hat{P}_{N_\nu}(A) = N_\nu^{-1} \sum_{i \in U_\nu} \mathbb{1}_{\{\mathbf{x}_i \in A\}}$$

each defined with respect to the finite population $U_\nu$.

Last we will define two functions of $N_\nu$ that will be used as rates of convergence. Let $\gamma(N_\nu)$ and $k(N_\nu)$ be given functions bounded above 0 for all $N_\nu > 0$ satisfying:

Rate Condition 1: $\gamma(N_\nu) \to \infty$,
Rate Condition 2: $N_\nu^{-1} k(N_\nu) \to 0$,
Rate Condition 3: $k(N_\nu)^{-1} \gamma(N_\nu) N_\nu^{1/2} \to 0$.

Note that Rate Conditions 1–3 require

$$N_\nu^{-1/2} \gamma(N_\nu) \to 0. \quad (4)$$

It is easy to find functions $k(N_\nu)$ and $\gamma(N_\nu)$ that satisfy Rate Conditions 1–3. For example, let $\alpha \in (1/2, 1)$ and $\epsilon \in (0, \alpha - 1/2)$. Then let $k(N_\nu) = N_\nu^\alpha$ and $\gamma(N_\nu) = \ln(N_\nu)$ or $N_\nu^{\alpha - \epsilon - 1/2}$.

For a given set $A \subset$ supp($\mathbf{X}$) the notation $\#_{N_\nu}(A) = \sum_{i \in U_\nu} \mathbb{1}_{\{\mathbf{x}_i \in A\}}$ will also be used throughout this article. The function $\#_{N_\nu}(A)$ represents the number of population elements that are contained in the set $A$.

To define the estimator of the function $h(\mathbf{x}) = E_\xi[Y|\mathbf{X} = \mathbf{x}]$ we will use the identity

$$E_\xi[Y|\mathbf{X} = \mathbf{x}] = \int_0^\infty \{1 - F(y|\mathbf{X} = \mathbf{x})\} dy$$

for $Y \geq 0$. The finite population estimator is given by

$$\hat{h}_{N_\nu}(\mathbf{x}) = \begin{cases} \int_0^{\gamma(N_\nu)} \{1 - \hat{F}_{N_\nu}(y|B^{(N_\nu)}(\mathbf{x}))\} dy, \\ \quad \text{if } \#_{N_\nu}(B^{N_\nu}(\mathbf{x})) > k(N_\nu) \\ 0, \quad \text{otherwise.} \end{cases} \quad (5)$$

By using the decomposition $Y = Y^+ - Y^-$ the definition of $h_{N_\nu}(\mathbf{x})$ can be made general.

The above estimator is a mean of values below the trimming cutoff of $\gamma(N_\nu)$ within a box containing at least $k(N_\nu)$ points. For sparse boxes [boxes containing fewer than $k(N_\nu)$ points] the estimator is set to zero.

*Theorem 1* (Gordon–Olsen 1980). Suppose the iid data $\{(Y_1, \mathbf{X}_1), \ldots, (Y_{N_\nu}, \mathbf{X}_{N_\nu})\}$ are from the superpopulation model $\xi$ with the properties that $\mathbf{X} \in \mathbb{R}^d$, and $E_\xi |Y|^p < \infty$. In addition assume, $k(N_\nu)$ and $\gamma(N_\nu)$ are two functions satisfying Rate Conditions 1–3. If

$$\hat{P}_{N_\nu}\{\mathbf{x}|\#(B^{N_\nu}(\mathbf{x})) > k(N_\nu)\} \to 1 \quad \text{with } \xi\text{-probability } 1, \quad (6)$$

and

$$\|Q^{N_\nu}\|_l^{\hat{F}_{N_\nu}} \to 0 \quad \text{and}$$
$$\|Q^{N_\nu}\|_l^{\hat{F}_{N_\nu}^-} \to_\xi 0 \quad \text{with } \xi\text{-probability } 1, \quad (7)$$

then

$$\lim_{N \to \infty} E_\xi\left[|h(\mathbf{X}) - \hat{h}_{N_\nu}(\mathbf{x})|^p\right] = 0. \quad (8)$$

Notice the trimming of this estimator vanishes if $\gamma(N_\nu)$ is sufficiently large compared to observed values of $Y$. However, Rate Condition 3 dictates that the faster $\gamma(N_\nu)$ grows the faster $k(N_\nu)$ must grow. This could cause more boxes to be too sparse, resulting in failing condition (6). Condition (6) requires that the proportion of boxes that are sufficiently dense as estimated from the empirical distribution has to go to 1. Condition (7) requires that the width of those boxes as measured by the empirical version of (2) and (3) shrink to zero. Note that these conditions are stated in terms of the finite population proportion $\hat{P}_{N_\nu}$ and the empirical norms $\|Q^{N_\nu}\|_l^{\hat{F}_{N_\nu}}$ and $\|Q^{N_\nu}\|_l^{\hat{F}_{N_\nu}^-}$.

## 3. COMPLEX SAMPLE DATA

Section 2 reviewed results contained in the work of Gordon and Olshen (1978, 1980) for the estimator $\hat{h}_{N_\nu}(\mathbf{x})$ using an iid sample of size $N_\nu$ from a superpopulation. These results established sufficient conditions for $L^p$ consistency of an estimator of $h(\mathbf{x})$ as $N_\nu$ grows without bound. This section considers an estimator of $h(\mathbf{x})$ based on recursive partitioning algorithms when the data are collected via a random sample with unequal probabilities of selection with replacement. For this end, we consider the performance of an estimator determined by the data from the sample of size $n_\nu$ in estimating $\hat{h}_{N_\nu}(\mathbf{x})$ as both $n_\nu, N_\nu \to \infty$.

Consider a population of elements indexed by the set $U_\nu = \{1, 2, \ldots, N_\nu\}$ where for every $i \in U_\nu$ there is a corresponding unit in the population $(Y_i, \mathbf{X}_i, \mathbf{Z}_i)$. We will again assume that the finite population is generated iid from a superpopulation model $\xi$. Here $Y_i$ and the vector $\mathbf{X}_i$ are the variables of interest. The vector $\mathbf{Z}_i$ represents known characteristics of the population elements that are used in the sample design but are not of direct interest to the analyst. For example, $\mathbf{Z}_i$ might contain a measure of size used to establish selection probabilities.

In contrast to the iid situation, in survey sampling it is often the case that some or all of the auxiliary variables are known for all units $i$ in the finite population and that this information could be used in designing the survey. In this article we examine the case in which $(Y_i, \mathbf{X}_i)$ is known only for the sampled elements $i \in S_\nu$, and only $\mathbf{Z}_i$ is known for each population unit, $i \in U_\nu$.

To examine asymptotic properties of an estimator we follow the framework of Isaki and Fuller (1982) and Robinson and Särndal (1983). Define a sequence of finite populations $\{U_\nu\}_{\nu=1}^\infty$, where $U_\nu \subset U_{\nu+1}$. The set $S_\nu \subset U_\nu$ represents a sample of size $n_\nu$ drawn from $U_\nu$. The probability of a particular sample $S_\nu$ being drawn among all possible samples from $U_\nu$ is called the sample design and is assumed known. Define $\delta_{\nu i} = 1$ if $i \in S_\nu$ and 0 otherwise. The probability of a particular unit $i$ being selected in the sample, $P(i \in S_\nu) = E_p(\delta_{\nu i})$, is denoted $\pi_{\nu i}$. The joint probability of two elements $i$ and $j$ being included in the sample, $P(i \in S_\nu \cap j \in S_\nu)$, is denoted $\pi_{\nu ij}$. Define $\tilde{N}_\nu = \sum_{i \in S_\nu} \pi_{\nu i}^{-1}$ as a weighted sample estimator of the population size.

In parallel with the definitions of $B_\nu^{N_\nu}(\mathbf{x})$ and $Q_\nu^{N_\nu}$ in Section 2, we define below the partition $Q^{n_\nu}$ and boxes $B^{n_\nu}(\mathbf{x})$ resulting from a partitioning of the $\mathbf{X}$-space using only the observed sample data. First we will need a few definitions.

Given a population of size $N_\nu$ and a box $B^{n_\nu}$ from a given partition $Q^{n_\nu}$, denote the number of population units in box $B^{n_\nu}$ as $\#_{N_\nu}(B^{n_\nu})$, and the number of units from sample $S_\nu$ contained in box $B^{n_\nu}$ as $\#_{n_\nu}(B^{n_\nu})$. We also define the sample based predictor

$$\tilde{\#}_{N_\nu}(B^{n_\nu}) = \sum_{i \in S_\nu} \pi_{\nu i}^{-1} \mathbb{1}_{\{\mathbf{x}_i \in B^{n_\nu}(\mathbf{x})\}}$$

of the number of population units in box $B^{n_\nu}$.

The partitioning algorithm we will use is based on one used by Gordon and Olshen (1978, 1980). The algorithm is first applied to the entire sample dataset and then recursively applied to each split subset.

Step 1. If the set contains more than $2k(n_\nu)$ sample elements go to Step 2; else stop.

Step 2. Among the auxiliary variables $X_j$, $j = 1, \ldots, d$ determine the split resulting in the largest decrease in the estimated mean squared error.

Step 3. If the minimizing split reduces the estimated mean squared error by at least $p\%$, then split the datasets according to this minimizing split. For each of the two resulting subsets, return to Step 1.

Step 4. Otherwise, split the set at the median value of the auxiliary variable $X_j$ that was least recently used by the algorithm to split. For each of the two resulting subsets, return to Step 1.

As was done for the finite population, define estimators of the empirical conditional distribution and the empirical probability function for $Y$ based on the sampled data. For a given box in the partition $B^{n_\nu}(\mathbf{x})$ define the probability-weighted empirical conditional distribution estimator

$$\tilde{F}_{n_\nu}(y|B^{n_\nu}(\mathbf{x})) = \left(\tilde{\#}_{N_\nu}(B^{n_\nu})\right)^{-1} \sum_{i \in S_\nu} \pi_{\nu i}^{-1} \mathbb{1}_{\{y_i \leq y\}} \mathbb{1}_{\{\mathbf{x}_i \in B^{n_\nu}(\mathbf{x})\}}$$

and the empirical probability estimator

$$\tilde{P}_{n_\nu}(B^{n_\nu}(\mathbf{x})) = (\tilde{N}_\nu)^{-1} \sum_{i \in S_\nu} \pi_{\nu i}^{-1} \mathbb{1}_{\{\mathbf{x}_i \in B^{n_\nu}(\mathbf{x})\}}.$$

### 3.1 Conditions

To prove consistency of an estimator based on the recursive partitioning of a sample with unequal probabilities of selection, we put some conditions on the sample design as well as the underlying population. The conditions are:

Condition 1

$$\lim_{\nu \to \infty} N_\nu^{-1} \sum_{i=1}^{N_\nu} Y_i^2 = \mu_2 < \infty \quad \text{with } \xi\text{-probability 1};$$

Condition 2

$$\limsup_{\nu \to \infty} \left( N_\nu \min_{i \in U_\nu} \pi_{\nu i} \right)^{-1} = O(n_\nu^{-1}) \quad \text{with } \xi\text{-probability 1};$$

Condition 3

$$\limsup_{\nu \to \infty} \max_{i,j \in U_\nu, i \neq j} \left| \frac{\pi_{\nu ij}}{\pi_{\nu i} \pi_{\nu j}} - 1 \right| = O(n_\nu^{-1}) \quad \text{with } \xi\text{-probability 1};$$

Condition 4

$$E_p[\delta_{\nu i} \delta_{\nu j} | Q^{n_\nu}] = \pi_{\nu ij} + O_p\big(n_\nu^{1/2} \gamma(n_\nu)^{-1} k(n_\nu)^{-1}\big)$$

$$\forall i, j \in U_\nu \text{ with } \xi\text{-probability 1}.$$

Condition 1 is a standard condition on the superpopulation while Conditions 2 and 3 are well-known conditions on the sample design that are easily satisfied by a survey design that is not highly clustered (see Robinson and Särndal 1983). Condition 4 is a condition concerning the finite population, the sample design, and the recursive partitioning algorithm. This condition requires that extreme values in the finite population selected into the sample have a diminishing effect on the recursive partitioning algorithm as the sample size contained in each box, $k(n_\nu)$, increases. See the article by John (1995) for a discussion of recursive partitioning methods that remove outliers. Notice that the factor $n_\nu^{1/2} \gamma(n_\nu)^{-1}$ acts to slow the required rate for this condition. The concept of diminishing effects here involves trade-offs among the rate functions $\gamma$ and $k$.

If we define the random variable

$$r_{\nu ij} = E_p[\delta_{\nu i} \delta_{\nu j} | Q^{n_\nu}] - \pi_{\nu ij},$$

then Condition 4 requires $\max_{i,j} |r_{\nu ij}| = O_p(n_\nu^{1/2} \gamma(n_\nu)^{-1} \times k(n_\nu)^{-1})$. Notice that every $r_{\nu ij}$ also satisfies $-1 \leq -\pi_{\nu ij} \leq r_{\nu ij} \leq 1 - \pi_{\nu ij} < 1$, with $\xi$-probability 1, so

$$E_p\left[ \max_{i,j} |r_{\nu ij}| \right] = O\big(n_\nu^{1/2} \gamma(n_\nu)^{-1} k(n_\nu)^{-1}\big). \tag{9}$$

### 3.2 Preliminary Results on Box Properties

Define $Y_\nu = \sum_{i \in U_\nu} Y_i$. The Hájek estimator of a population mean $\bar{Y}_\nu = N_\nu^{-1} \sum_{i \in U_\nu} Y_i = N_\nu^{-1} \hat{Y}_\nu$ is defined as the ratio estimator

$$\hat{\bar{Y}}_\nu = \left( \sum_{i \in U_\nu} \pi_{\nu i}^{-1} \delta_{\nu i} \right)^{-1} \sum_{i \in U_\nu} \pi_{\nu i}^{-1} \delta_{\nu i} y_i = \tilde{N}_\nu^{-1} \hat{Y}_\nu.$$

Rather than prove that an estimator defined on a recursive partitioning created from the algorithm applied to sample data converges to a partitioning created from the algorithm applied to the population data, we proceed by showing the following.

- The partition $Q^{n_\nu}$, created by applying the algorithm to the sample data, satisfies the conditions corresponding to expressions (6) and (7).
- Consequently, the estimator $\hat{h}^*_{N_\nu}(\mathbf{x})$, defined using the *finite population data* on the *sample based* partition $Q^{n_\nu}$, is consistent as an estimator of the superpopulation quantity $h(\mathbf{x})$.
- The estimator defined using the sample is ADU and ADC as an estimator for $\hat{h}^*_{N_\nu}(\mathbf{x})$ defined using the population values on the partition.

This will prove that the estimator defined using the sample is ADU and ADC as an estimator of the superpopulation quantity $h(\mathbf{x})$.

*Lemma 1.* Assume the sample design satisfies Conditions 1 through 4, and $k$ and $\gamma$ are two functions satisfying Rate Conditions 1 through 3. If with $\xi$-probability 1,

$$\lim_{\nu \to \infty} \tilde{P}_{n_\nu}\big[\mathbf{x} | k(n_\nu)^{-1} \#_{n_\nu}(B^{n_\nu}(\mathbf{x})) \geq 1\big] = 1,$$

then

$$\lim_{\nu \to \infty} \hat{P}_{N_\nu}\big[\mathbf{x} | k(n_\nu)^{-1} \#_{n_\nu}(B^{n_\nu}(\mathbf{x})) \geq 1\big] = 1$$

with $\xi$-probability 1.

See Appendix A for proof.

Next we define a sample based estimator for the $l$-norm of partition $Q^{n_\nu}$ relative to $\hat{F}_{N_\nu}$. This is done by using $\hat{F}_{N_\nu}$ and $\tilde{F}_{N_\nu}$, respectively, and their corresponding empirical probability measures in place of $F$ and $P$ in Equation (2).

*Lemma 2.* Under Conditions 1 through 4, given two functions $k$ and $\gamma$ satisfying Rate Conditions 1 through 3, then for all $l = 1, \ldots, d$

$$\|Q^{n_\nu}\|_l^{\tilde{F}_{n_\nu}} \to_p \|Q^{n_\nu}\|_l^{\hat{F}_{N_\nu}} \quad \text{and} \quad \|Q^{n_\nu}\|_l^{\tilde{F}_{N_\nu}^-} \to_p \|Q^{n_\nu}\|_l^{\hat{F}_{N_\nu}^-},$$

with $\xi$-probability 1 as $\nu \to \infty$.

See Appendix A for proof.

### 3.3 The Mean Estimator

Next is the main result of this article. In addition to the previous conditions on the sample design, this result requires the added condition on the sampling fraction:

Condition 5

$$f^{-1} = N_\nu/n_\nu = O\big(n_\nu^{1/2} \gamma(n_\nu)^{-1}\big).$$

This condition allows for designs with a finite sampling fraction $f > 0$, as well as a sampling fraction that shrinks asymptotically to zero.

*Proposition 1.* Given a finite population $\{(Y_1, \mathbf{X}_1), \ldots, (Y_{N_\nu}, \mathbf{X}_{N_\nu})\}$ generated iid from the superpopulation model $\xi$ and a sample $S_\nu$, selected from this population satisfying Conditions 1 through 5. Let $Q^{n_\nu}$ be the collection of boxes created from the algorithm, which partition the finite population and define

$$\tilde{h}_{n_\nu}(\mathbf{x}) = \begin{cases} \int_0^{\gamma(n_\nu)} \{1 - \tilde{F}_{n_\nu}(y | B^{n_\nu}(\mathbf{x}))\} \, dy, \\ \quad \text{if } \#_{n_\nu}(B^{n_\nu}) > k(n_\nu) \\ 0, \quad \text{otherwise.} \end{cases} \tag{10}$$

Assume the two functions $k$ and $\gamma$ satisfy Rate Conditions 1 through 3.

If with $\xi$-probability 1,

$$\tilde{P}\big(\mathbf{x}|k(n_\nu)^{-1}\#_{n_\nu}(B^{n_\nu}(\mathbf{x})) \geq 1\big) \to_p 1 \qquad (11)$$

and

$$\begin{aligned}\|Q^{n_\nu}\|_l^{\tilde{F}_{n_\nu}} &\to_p 0 \qquad \text{and} \\ \|Q^{n_\nu}\|_l^{\tilde{F}_{n_\nu}^-} &\to_p 0 \quad \text{for all } l = 1, \ldots, d,\end{aligned} \qquad (12)$$

then

$$\lim_{\nu \to \infty} E_{\xi p}\big[|\tilde{h}_{n_\nu}(\mathbf{x}) - h(\mathbf{x})|^2\big] = 0.$$

*Proof.* Define the finite population estimator of $h(\mathbf{x})$ using the resulting partition from the recursive partitioning algorithm used on the sample data as

$$\hat{h}^*_{N_\nu}(\mathbf{x}) = \begin{cases} \int_0^{\gamma(n_\nu)} \{1 - \hat{F}_{N_\nu}(y|B^{n_\nu}(\mathbf{x}))\}\,dy, \\ \qquad \text{if } \#_{N_\nu}(B^{n_\nu}(\mathbf{x})) > k(n_\nu) \\ 0, \quad \text{otherwise.} \end{cases} \qquad (13)$$

Notice that $\hat{h}^*_{N_\nu}(\mathbf{x})$ is a finite population quantity defined on the sample based partition. Since the analyst only observes the sample, this is a hypothetical quantity used here as a device to prove consistency of the sample estimator $\tilde{h}_{n_\nu}(\mathbf{x})$.

We will show that the quantity

$$\begin{aligned}&E_{\xi p}\big[|\tilde{h}_{n_\nu}(\mathbf{x}) - h(\mathbf{x})|^2\big] \\ &= E_{\xi p}\big[|\tilde{h}_{n_\nu}(\mathbf{x}) - \hat{h}^*_{N_\nu}(\mathbf{x})|^2\big] \\ &\quad + 2E_{\xi p}\big[|\tilde{h}_{n_\nu}(\mathbf{x}) - \hat{h}^*_{N_\nu}(\mathbf{x})||\hat{h}^*_{N_\nu}(\mathbf{x}) - h(\mathbf{x})|\big] \\ &\quad + E_{\xi p}\big[|\hat{h}^*_{N_\nu}(\mathbf{x}) - h(\mathbf{x})|^2\big]\end{aligned}$$

goes to zero as $n_\nu$ increases.

Let $\mathcal{A}$ be the event that the samples are chosen such that the assumptions (11) and (12) imply that

$$\lim_{\nu \to \infty} \hat{P}\big[x|k(n_\nu)^{-1}\#_{n_\nu}(B^{n_\nu}(\mathbf{x})) \geq 1\big] = 1$$

and

$$\lim_{\nu \to \infty} \|Q^{n_\nu}\|_l^{\hat{F}_{N_\nu}} = 0 \qquad \text{and} \qquad \lim_{\nu \to \infty} \|Q^{n_\nu}\|_l^{\hat{F}_{N_\nu}^-} = 0$$

with $\xi$-probability 1. Then

$$E_{\xi p}\big[|\hat{h}^*_{N_\nu}(\mathbf{x}) - h(\mathbf{x})|^2\big] = E_\xi\big[|\hat{h}^*_{N_\nu}(\mathbf{x}) - h(\mathbf{x})|^2|\mathcal{A}\big]P_p(\mathcal{A}) \\ + E_\xi\big[|\hat{h}^*_{N_\nu}(\mathbf{x}) - h(\mathbf{x})|^2|\mathcal{A}'\big]P_p(\mathcal{A}')$$

where $\mathcal{A}'$ is the complement of event $\mathcal{A}$.

The quantity $E_\xi[|\hat{h}^*_{N_\nu}(\mathbf{x}) - h(\mathbf{x})|^2|\mathcal{A}'] \leq E_\xi[|\hat{h}^*_{N_\nu}(\mathbf{x})|^2|\mathcal{A}'] + E_\xi[|h(\mathbf{x})|^2|\mathcal{A}']$, where the estimator $\hat{h}^*_{N_\nu}(\mathbf{x})$ is a mean of $Y_i$'s and $h(x) = Y_i + \epsilon_i$. By Condition 1 $E_\xi[Y_i^2] < \infty$ with $\xi$-probability 1; therefore the quantity $E_\xi[Y_i^2|\mathcal{A}']$ is bounded because $\mathcal{A}'$ is a event determined by the sample. This implies that the expectation $E_\xi[|\hat{h}^*_{N_\nu}(\mathbf{x}) - h(\mathbf{x})|^2|\mathcal{A}']$ is bounded.

On the other hand, if the event $\mathcal{A}$ occurs, $E_\xi[|\hat{h}^*_{N_\nu}(\mathbf{x}) - h(\mathbf{x})|^2|\mathcal{A})$ converges to zero by Theorem 1. Since we have shown in Lemma 1 and Lemma 2 that $P_p(\mathcal{A}) \to 1$, it follows that

$$\lim_{\nu \to \infty} E_\xi\big[|\hat{h}^*_{N_\nu}(\mathbf{x}) - h(\mathbf{x})|^2\big] = 0.$$

We proceed to show that

$$\lim_{\nu \to \infty} E_p\big[|\tilde{h}_{n_\nu}(\mathbf{x}) - \hat{h}^*_{N_\nu}(\mathbf{x})|^2\big] = 0$$

with $\xi$-probability 1.

The set of $\mathbf{x}$ such that $\#_{N_\nu}(B^{n_\nu}(\mathbf{x})) > k(n_\nu)$ is a set with $\xi$-measure that converges to 1 as $n_\nu$ increases. Thus, we will only consider $\mathbf{x}$ satisfying this condition. For such an $\mathbf{x}$, the difference $\tilde{h}_{n_\nu}(\mathbf{x}) - \hat{h}^*_{N_\nu}(\mathbf{x})$ can be written $D_1 + D_2$, where

$$D_1 = \int_0^{\gamma(n_\nu)} \#_{N_\nu}(B^{n_\nu}(\mathbf{x}))^{-1} \\ \times \sum_{i \in U_\nu} \mathbb{1}_{\{y_i \leq y\}} \mathbb{1}_{\{\mathbf{x}_i \in B^{(n_\nu)}(\mathbf{x})\}} (1 - \pi_{\nu i}^{-1}\delta_{\nu i})\,dy$$

and

$$D_2 = \big[\#_{N_\nu}(B^{n_\nu}(\mathbf{x}))^{-1} - \tilde{\#}_{N_\nu}(B^{n_\nu}(\mathbf{x}))^{-1}\big] \\ \times \int_0^{\gamma(n_\nu)} \sum_{i \in U_\nu} \pi_{\nu i}^{-1}\delta_{\nu i}\mathbb{1}_{\{y_i \leq y\}}\mathbb{1}_{\{\mathbf{x}_i \in B^{n_\nu}(\mathbf{x})\}}\,dy.$$

In addition,

$$|D_1| \leq \gamma(n_\nu)\#_{N_\nu}(B^{n_\nu}(\mathbf{x}))^{-1}\left|\sum_{i \in U_\nu}(1 - \pi_{\nu i}^{-1}\delta_{\nu i})\mathbb{1}_{\{\mathbf{x}_i \in B^{(n_\nu)}(\mathbf{x})\}}\right|$$

and

$$|D_2| \leq \#_{N_\nu}(B^{n_\nu}(\mathbf{x}))^{-1}\big|\#_{N_\nu}(B^{n_\nu}(\mathbf{x})) - \tilde{\#}_{N_\nu}(B^{n_\nu}(\mathbf{x}))\big| \\ \times \int_0^{\gamma(n_\nu)} \tilde{F}_{n_\nu}(y|B^{n_\nu}(\mathbf{x}))\,dy.$$

Each of $|D_1|$ and $|D_2|$ is

$$\leq \gamma(n_\nu)\#_{N_\nu}(B^{n_\nu}(\mathbf{x}))^{-1}\left|\sum_{i \in U_\nu}(1 - \pi_{\nu i}^{-1}\delta_{\nu i})\mathbb{1}_{\{\mathbf{x}_i \in B^{(n_\nu)}(\mathbf{x})\}}\right|$$

with $\xi$-probability 1.

Therefore,

$$\begin{aligned}&E_p\big[\{\tilde{h}_{n_\nu}(\mathbf{x}) - \hat{h}^*_{N_\nu}(\mathbf{x})\}^2\big] \\ &\leq 4\gamma^2(n_\nu)E_p\bigg[\#_{N_\nu}(B^{n_\nu}(\mathbf{x}))^{-2}\sum_{i,j \in U_\nu}(1 - \pi_{\nu i}^{-1}\delta_{\nu i}) \\ &\qquad \times (1 - \pi_{\nu j}^{-1}\delta_{\nu j})\mathbb{1}_{\{\mathbf{x}_i, \mathbf{x}_j \in B^{(n_\nu)}(\mathbf{x})\}}\bigg] \\ &= 4\gamma^2(n_\nu)E_p\bigg[E_p\bigg\{\#_{N_\nu}(B^{n_\nu}(\mathbf{x}))^{-2}\sum_{i,j \in U_\nu}(1 - \pi_{\nu i}^{-1}\delta_{\nu i}) \\ &\qquad \times (1 - \pi_{\nu j}^{-1}\delta_{\nu j})\mathbb{1}_{\{\mathbf{x}_i, \mathbf{x}_j \in B^{(n_\nu)}(\mathbf{x})\}}\bigg|Q^{n_\nu}\bigg\}\bigg] \\ &= 4\gamma^2(n_\nu)E_p\bigg[\#_{N_\nu}(B^{n_\nu}(\mathbf{x}))^{-2}\sum_{i,j \in U_\nu}\mathbb{1}_{\{\mathbf{x}_i, \mathbf{x}_j \in B^{(n_\nu)}(\mathbf{x})\}} \\ &\qquad \times E_p\{(1 - \pi_{\nu i}^{-1}\delta_{\nu i})(1 - \pi_{\nu j}^{-1}\delta_{\nu j})|Q^{n_\nu}\}\bigg]\end{aligned}$$

$$= 4\gamma^2(n_\nu) E_p \left[ \#_{N_\nu}(B^{n_\nu}(\mathbf{x}))^{-2} \sum_{i \in U_\nu} \mathbb{1}_{\{\mathbf{x}_i \in B^{(n_\nu)}(\mathbf{x})\}} \right.$$

$$\left. \times \{\pi_{\nu i}^{-1} - 1 + r_{\nu ij}\} \right]$$

$$+ 4\gamma^2(n_\nu) E_p \left[ \#_{N_\nu}(B^{n_\nu}(\mathbf{x}))^{-2} \sum_{i \neq j \in U_\nu} \sum \mathbb{1}_{\{\mathbf{x}_i, \mathbf{x}_j \in B^{(n_\nu)}(\mathbf{x})\}} \right.$$

$$\left. \times \left\{ \frac{\pi_{\nu ij}}{\pi_{\nu i} \pi_{\nu j}} - 1 + r_{\nu ij} \right\} \right].$$

Since $\#_{N_\nu}(B^{n_\nu}(\mathbf{x})) \geq k(n_\nu)$, the first term is

$$\leq 4\gamma^2(n_\nu) k(n_\nu)^{-1}$$

$$\times \left\{ N_\nu \left( N_\nu \min_{i \in U_\nu} \pi_{\nu i} \right)^{-1} + E_p \left[ \max_{i,j} |r_{\nu ij}| \right] \right\}. \quad (14)$$

By Condition 2 and Equation (9), Equation (14) is $O((N_\nu/n_\nu)\gamma^2(n_\nu)k(n_\nu)^{-1})$, with $\xi$-probability 1. By Condition 5, this quantity is $O(n_\nu^{1/2}\gamma(n_\nu)k(n_\nu)^{-1})$.

Likewise, the second term is

$$\leq 4\gamma^2(n_\nu) \left\{ \max_{i,j \in U_\nu, i \neq j} \left| \frac{\pi_{\nu ij}}{\pi_{\nu i}\pi_{\nu j}} - 1 \right| + E_p \left[ \max_{i,j} |r_{\nu ij}| \right] \right\}$$

which, by Condition 2 and Equation (9), is $O(n_\nu^{1/2}\gamma(n_\nu) \times k(n_\nu)^{-1})$ with $\xi$-probability 1.

Therefore, $E_p[\{\tilde{h}_{n_\nu}(\mathbf{x}) - \hat{h}^*_{N_\nu}(\mathbf{x})\}^2] = O(n_\nu^{1/2}\gamma(n_\nu)k(n_\nu)^{-1})$ with $\xi$-probability 1, which, by Rate Condition 3, goes to zero.

## 4. APPLICATION: ESTABLISHMENT PAYROLL DATA

The average wage per employee paid by an establishment depends a great deal on location of the establishment and the type of industry in which the establishment is engaged. Among the 11 different North American Industry Classification System (NAICS) super sectors one of the lowest paying is the leisure and hospitality industry class (2-digit NAICS codes 71 and 72). In fact, the average total wage per employee paid by an establishment with the leisure and hospitality classification in the second quarter of 2005 is $8827. Excluding the establishments in the largest metropolitan statistical area (MSA), the MSA category with the largest average wage per employee, this average is $7685. Among these establishments the average is only $3596 for establishments in the leisure and hospitality industry class while $8207 for the rest. Even among establishments in this industry classification the average wage per employee can vary greatly depending on the characteristics of an establishment.

The OES is a semi-annual establishment survey which measures a number of establishment level characteristic variables. In this study, we use four establishment characteristics to model the average wage per employee. The four characteristics of interest are: (1) *size*, the size of the establishment measured by the number of employees employed at the establishment, (2) *age*, the age of the establishment's parent firm measured in years since it has been open, (3) *msa*, the size class of the MSA, and (4) *count*, the number of other establishments owned by the parent firm in different states. The last characteristic is often used as a proxy for the complexity of the corporate structure of the establishment's parent company firm.

The measure of average wages paid per employee for an establishment is derived by taking the reported payroll for the second quarter of 2005 according to the Quarterly Census of Employment and Wages (QCEW) divided by the reported number of employees for the selected establishment. The average wage per employee of small establishments can have extreme behavior. For example, some family owned and operated establishments may not pay anything to some or all of their employees. Consequently we limit our analysis only to establishments with reported wages above zero and at least 20 employees. We also exclude establishments in the largest MSA class (those with more than one million people). Figure 1 shows the resulting regression trees based on the 7112 sample units from the May 2006 OES linked with their payroll data. The goal was to model the association between an establishment's characteristics and the average wage per employee paid by that establishment for both the unadjusted and weight-adjusted algorithm.

Concentrating on only the first two splits of the tree models (Figure 2), we see a qualitative similarity between the two models. Using just these two splits, the establishments are divided into three categories by both algorithms: large firms; old small firms; and young small firms. Differences between the weighted and unweighted model results include the definitional boundaries for these categories, as well as the estimate of the average wages paid for each category. For example, establishments are singled out by the unadjusted algorithm as large if they have at least 148 employees, whereas the weight-adjusted algorithm defines large as having at least 83 employees. The difference in cutting-point in this case can be attributed to the difference between the weighted and unweighted versions of the empirical distribution functions (edf). For the unweighted edf the point 148 employees is the 95% quantile while the point 83 employees represents only the 84% quantile. For the weighted version 83 employees is close to the 95% quantile (93%).

Cut-points on the variable *age* were relatively close: 34.41 years for the unadjusted algorithm and 33.75 years for the weight-adjusted algorithm. In addition, the two methods produced different estimates of average wage per employee for each category. The unadjusted model estimates an average wage which is higher than that of the weight-adjusted model for all three categories of establishments.

### 4.1 Simulation

To explore potential effects of ignoring the weights when applying a recursive partitioning algorithm we provide a simulation that uses this dataset to compare the two options. The dataset is now considered the finite population of size $N_\nu = 7112$ that we intend to model. We will compare the accuracy of the regression tree model ignoring the weights to the model that incorporates the sampling weights using the finite population estimator $\hat{h}_{N_\nu}$ as a proxy for $h(x)$. For this end, we draw 1000 repeated samples from this finite population of size $n_\nu$ according to one of two sample designs and compare the models obtained by the two recursive partitioning methods on each sample to the true finite population. The first design uses the relative probabilities of selection for each sampled unit used to draw the original OES sample. How this was done is explained in detail below.

Both sample designs are single stage probability proportional to size (pps) designs. For the first design (denoted *OES Design*)

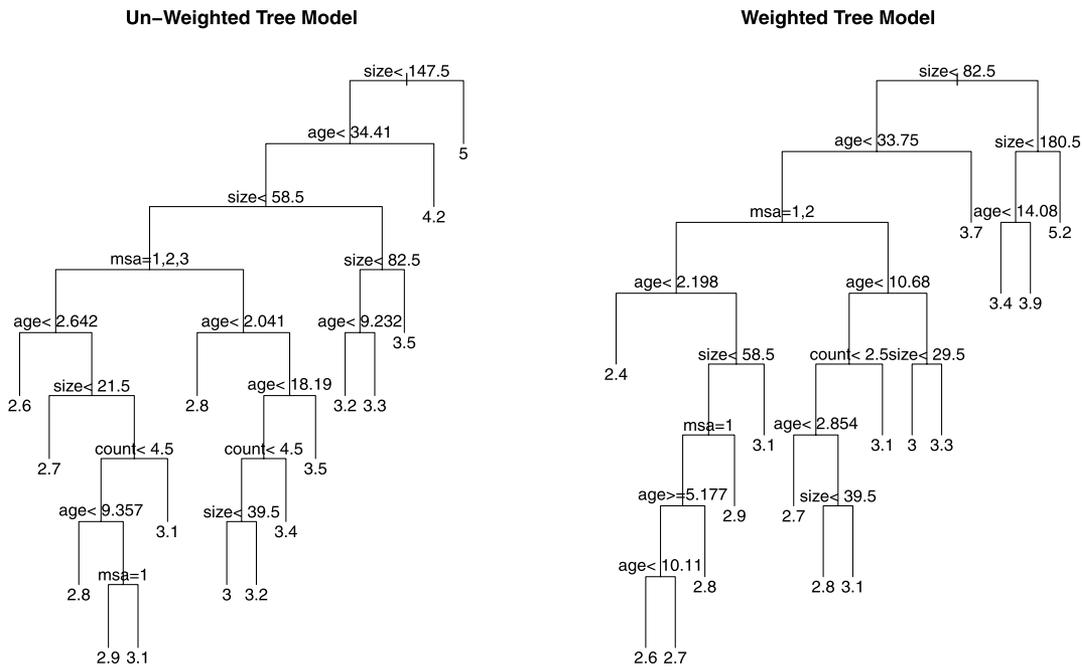

Figure 1. Shows both trees modeling average wage per employee (in 000's) paid by an establishment with at least 20 employees in the Leisure and Hospitality category located outside of a large metropolitan area. The algorithm used the 7112 units from the May 2006 OES sample of establishments linked with reported wages for the second quarter of 2005.

we use the inverses of the original OES sampling weights as the measure of size variables. In the second design (denoted *PPS Design*) we use the number of employees at an establishment. Under each of the two designs every element *i* in the population has a positive probability $\pi_i$ of being selected in the sample. Table 1 provides characteristics of the two designs used in the simulation. From this table we see that the sampling schemes are very similar with the exception of the presence of certainty units in every sample in the second design.

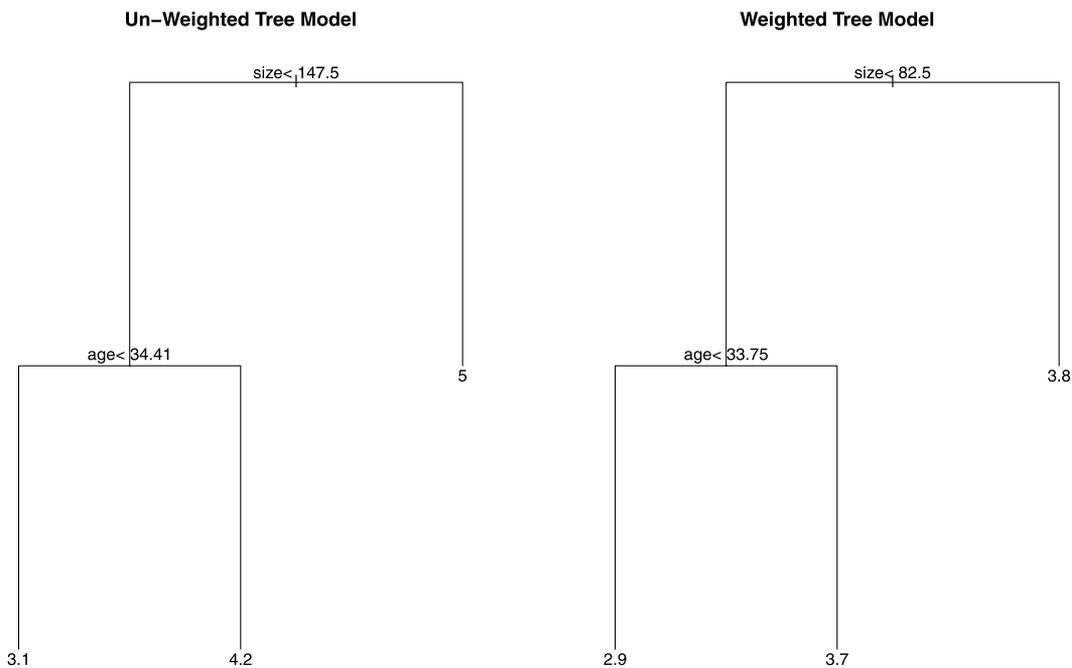

Figure 2. Shows both trees modeling average wage per employee (in 000's) paid by an establishment with at least 20 employees in the Leisure and Hospitality category located outside of a large metropolitan area. These trees resulted from pruning the full trees to include only the first two splits. The tree on the left results from ignoring the complex sample design and the sample weights while the tree on the right is the result of the procedure incorporating the sample weights. The recursive partitioning algorithm identifies size of the establishment and age of the firm as among the most influential characteristics in determining average wage per employee paid by an establishment. The tree splits the smaller values to the left. For instance, the algorithm identifies the small and young establishments as paying the lowest average wage per employee among these types of establishments.

Table 1. This table displays key characteristics of the two sample designs used to select the repeated samples in the simulations. The characteristics displayed are sample size, the number of certainty units (CUs), the minimum probability of selection for a single element, the maximum probability of selection for a single element, the coefficient of variation CV of the probabilities of selection, and the correlation of the probabilities of selection with the variable of interest, $y_i$, the average wage per employee

| Design | $n_\nu$ | CUs | min $\pi_\nu$ | max $\pi_\nu$ | CV($\pi_\nu$) | Cor($y_i, \pi_{\nu i}$) |
|---|---|---|---|---|---|---|
| OES Design | 100 | 0 | 0.001736 | 0.106153 | 1.103350 | 0.226363 |
|  | 200 | 0 | 0.003471 | 0.212306 | 1.103350 | 0.226363 |
|  | 400 | 0 | 0.006943 | 0.424611 | 1.103350 | 0.226363 |
|  | 800 | 0 | 0.013885 | 0.849222 | 1.103350 | 0.226363 |
|  | 1600 | 158 | 0.029821 | 1.000000 | 0.799255 | 0.248926 |
| PPS Design | 100 | 2 | 0.004286 | 1.000000 | 2.178212 | 0.163341 |
|  | 200 | 5 | 0.008694 | 1.000000 | 1.865246 | 0.177456 |
|  | 400 | 23 | 0.017931 | 1.000000 | 1.482854 | 0.200450 |
|  | 800 | 64 | 0.037626 | 1.000000 | 1.123322 | 0.223481 |
|  | 1600 | 172 | 0.080510 | 1.000000 | 0.817056 | 0.225856 |

In order to compare the two methods we define some statistics that we calculate for each sample. Let $U$ be the set of integers $\{1, \ldots, 7112\}$ and let $T$ be the regression tree model obtained by applying the recursive partitioning algorithm on the population elements. Denote the regression tree $T$ estimated value of average wage per employee paid $y(\mathbf{x}_i)$ by establishment $i \in U$ given the establishment's characteristic variables $\mathbf{x}_i$ by $T(\mathbf{x}_i)$. For each sample $S$ of population elements we build two regression trees using the recursive partitioning algorithms. The estimator $\tilde{t}$ uses the method proposed above to account for the sampling weights and the other $\hat{t}$ ignores the sampling weights. To compare the two resulting regression tree models we consider the mean error and the mean squared error with respect to the population model $T$. For each sample $S$ the mean error $N^{-1} \sum_{i \in U} \{\tilde{t}(\mathbf{x}_i) - T(\mathbf{x}_i)\}$ and the mean squared error $N^{-1} \sum_{i \in U} \{\tilde{t}(\mathbf{x}_i) - T(\mathbf{x}_i)\}^2$ for the regression model $\tilde{t}$ are calculated. The analogous quantities are produced for $\hat{t}$ as well.

Figure 3 demonstrates the behavior of the mean error and the square root of mean squared error for $\tilde{t}$ and $\hat{t}$ as the sample size increases. From this set of graphs it is clear that the weighted estimator performs much better than the estimator which ignores the design in terms of overall efficiency. More importantly, the weight-adjusted method appears to remove a substantial amount of bias from the estimate that would be present if an unadjusted method were used.

## 5. DISCUSSION

This article establishes sufficient conditions on the population distribution, survey design, and recursive partitioning algorithm that guarantee asymptotic design consistency of regression trees as an estimator for the conditional mean of the population. Consistency of the weighted estimator is demonstrated through an empirical investigation and simulation study of the methods applied to data based on the 2006 OES establishment data linked with 2005 second quarter QCEW payroll data. This investigation gives strong evidence that ignoring the complex design of the sample when using recursive partitioning techniques may have severe negative consequences. It may be of interest to note that empirical investigation of the techniques on other simulated models, not discussed in this article, showed similar results.

The algorithms described above are based completely on observed values from the sample as are the conditions needed to establish consistency. It may be possible to incorporate known totals for auxiliary variables into the algorithm to gain improved efficiency. This is an area we hope to investigate further in the future.

Further, all the results in this article are asymptotic under a superpopulation model; nothing in this article discusses properties for a single finite population. In particular, it would be desirable to have diagnostics to detect an extreme sample or finite population that would lead to misleading results. Alternatively, a procedure and conditions under which the estimator has almost sure convergence under the superpopulation model may be useful. In addition, determining sample designs that optimize recursive partition algorithms for a given finite population is an area for future study.

The asymptotic results in this article, like those of Gordon and Olshen (1978, 1980), deal only with building a tree model. When practitioners use this methodology they usually incorporate a pruning algorithm to complete the model selection process; see the book by Breiman et al. (1984). Pruning algorithms on data from a complex sample are therefore another possible area for future research. This article offers a beginning to providing theoretical justification for using recursive partitioning algorithms on survey data. However, there is clearly much research to be done in this area.

## APPENDIX A: PROOFS AND ADDITIONAL LEMMAS

*Lemma A.1.* Assume Conditions 1, 2, and 3. Then the Hájek estimator, $\hat{\bar{Y}}_\nu$, of a population mean is asymptotically design unbiased (ADU)

$$\lim_{\nu \to \infty} E_p[\hat{\bar{Y}}_\nu - \bar{Y}_\nu] = 0$$

with $\xi$-probability 1, and asymptotically design consistent (ADC)

$$\lim_{\nu \to \infty} P_p(|\hat{\bar{Y}}_\nu - \bar{Y}_\nu| > \epsilon) = 0$$

with $\xi$-probability 1, for any $\epsilon > 0$, where $E_p$ and $P_p$ denote expectation and probability, respectively, with respect to the sample design.

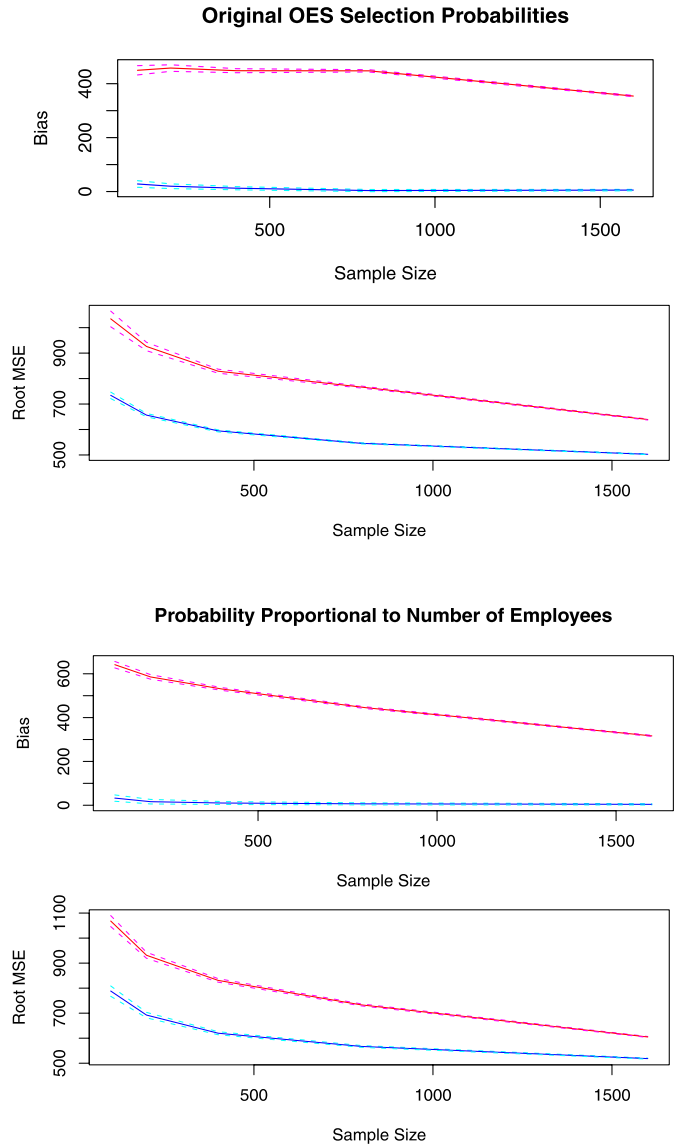

Figure 3. The solid lines represent the empirical bias (the average of the mean error calculations) and the square root of the empirical mean squared error (RMSE) based on 1000 simulations for the unweighted estimator (red—top) and the weighted estimator (blue—bottom). The dotted lines are ±1.96 the standard errors of the 1000 calculations for each sample size $n_\nu$. The online version of this figure is in color.

*Proof.* The Horvitz–Thompson estimator of a population mean is defined $N_\nu^{-1}\hat{Y}_\nu$. This estimator (Horvitz and Thompson 1952) is ADC as an estimator of the population mean $\bar{Y}$ under the above conditions (see Robinson and Särndal 1983, thm. 1). In fact $N_\nu^{-1}\hat{Y}_\nu = \bar{Y}_\nu + O_p(n_\nu^{-1/2})$ (see Robinson 1982; Mukhopadhyay 2006). This is also true for the special case $Y_i = 1$ for all $i$, $N_\nu^{-1}\tilde{N}_\nu = 1 + O_p(n_\nu^{-1/2})$.

The Hájek estimator can be expressed as the function of these two Horvitz–Thompson estimators $\hat{\bar{Y}}_\nu = f(\hat{Y}_\nu N_\nu^{-1}, \tilde{N}_\nu N_\nu^{-1})$ where $f(y, x) = y/x$. Since $f$ has a continuous derivative at $x = N_\nu > 0$ and $\tilde{N}_\nu \geq 1$ for all $\nu$, $\hat{\bar{Y}}_\nu = \bar{Y}_\nu / 1 + O_p(n_\nu^{-1/2})$ by the vector form of corollary 5.1.5 in the book by Fuller (1976).

*Lemma A.2.* Let $R_{n_\nu i} \geq 0$ be a random variable that depends on the sample, but is conditionally independent of the sample given $Q^{n_\nu}$ and $N_\nu^{-1}\sum_{i \in U_\nu} R_{n_\nu i}^2 \leq M < \infty$ for every $\nu$. Assume Conditions 1 through 4; then

$$\lim_{\nu \to \infty} E_p\left[\left\{N_\nu^{-1}\sum_{i \in U_\nu}(\delta_{\nu i}\pi_{\nu i}^{-1} - 1)R_{n_\nu i}\right\}^2\right] = 0 \quad (A.1)$$

with $\xi$-probability 1, where $E_p$ denotes expectation with respect to the sample design.

For example, $R_{n_\nu i}$ can be a deterministic function of the sample driven partition such as $R_{n_\nu i} = \mathbb{1}_{\{\mathbf{x}_i \in B^{n_\nu}(\mathbf{x})\}}$, for a predetermined $x$.

*Proof of Lemma A.2.* The quantity

$$E_p\left[\left\{N_\nu^{-1}\sum_{i \in U_\nu}(\delta_{\nu i}\pi_{\nu i}^{-1} - 1)R_{n_\nu i}\right\}^2\right]$$

$$= E_p\left[E_p\left[\left\{N_\nu^{-1}\sum_{i \in U_\nu}(\delta_{\nu i}\pi_{\nu i}^{-1} - 1)R_{n_\nu i}\right\}^2 \bigg| Q^{n_\nu}\right]\right]$$

$$= E_p\left[N_\nu^{-2}\sum_{i \in U_\nu}\{\pi_{\nu i}^{-1} - 1 + r_{\nu ij}\}R_{n_\nu i}^2\right]$$

$$+ E_p\left[N_\nu^{-2}\sum_{i \neq j \in U_\nu}\left\{\frac{\pi_{\nu ij}}{\pi_{\nu i}\pi_{\nu j}} - 1 + r_{\nu ij}\right\}R_{n_\nu i}R_{n_\nu j}\right]$$

$$\leq E_p\left[\left\{\left(N_\nu \min_{i \in U_\nu}\pi_{\nu i}\right)^{-1} + \max_{i,j}|r_{\nu ij}|\right\}N_\nu^{-1}\sum_{i \in U_\nu}R_{n_\nu i}^2\right]$$

$$+ E_p\left[\left\{\max_{i,j \in U_\nu, i \neq j}\left|\frac{\pi_{\nu ij}}{\pi_{\nu i}\pi_{\nu j}} - 1\right| + \max_{i,j}|r_{\nu ij}|\right\}\right.$$

$$\left.\times N_\nu^{-2}\sum_{i \neq j \in U_\nu} R_{n_\nu i}R_{n_\nu j}\right].$$

Since

$$N_\nu^{-2}\sum_{i \neq j \in U_\nu} R_{n_\nu i}R_{n_\nu j} \leq N_\nu^{-1}\sum_{i \in U_\nu}R_{n_\nu i}^2 < M$$

uniformly, the quantity

$$\leq \left\{\left(N_\nu \min_{i \in U_\nu}\pi_{\nu i}\right)^{-1} + E_p\left[\max_{i,j}|r_{\nu ij}|\right]\right\}M$$

$$+ \left\{\max_{i,j \in U_\nu, i \neq j}\left|\frac{\pi_{\nu ij}}{\pi_{\nu i}\pi_{\nu j}} - 1\right| + E_p\left[\max_{i,j}|r_{\nu ij}|\right]\right\}M,$$

which is $O(n_\nu^{1/2}\gamma(n_\nu)^{-1}k(n_\nu)^{-1})$ by Condition 2, Condition 3, and Equation (9).

Proof of Lemma 1

Let $A_\nu = \{x | k(n_\nu)^{-1}\#_{n_\nu}(B^{n_\nu}(\mathbf{x})) \geq 1\}$. The difference

$$|\tilde{P}(x|\mathbf{x} \in A_\nu) - \hat{P}(x|\mathbf{x} \in A_\nu)|$$

$$= \left|\tilde{N}_\nu^{-1}\sum_{i \in U_\nu}\pi_{\nu i}^{-1}\delta_{\nu i}\mathbb{1}_{\{\mathbf{x}_i \in A_\nu\}} - N_\nu^{-1}\sum_{i \in U_\nu}\mathbb{1}_{\{\mathbf{x}_i \in A_\nu\}}\right|$$

$$\leq \left|\tilde{N}_\nu^{-1}\sum_{i \in U_\nu}\pi_{\nu i}^{-1}\delta_{\nu i}\mathbb{1}_{\{\mathbf{x}_i \in A_\nu\}} - N_\nu^{-1}\sum_{i \in U_\nu}\pi_{\nu i}^{-1}\delta_{\nu i}\mathbb{1}_{\{\mathbf{x}_i \in A_\nu\}}\right|$$

$$+ N_\nu^{-1}\left|\sum_{i \in U_\nu}\pi_{\nu i}^{-1}\delta_{\nu i}\mathbb{1}_{\{\mathbf{x}_i \in A_\nu\}} - \sum_{i \in U_\nu}\mathbb{1}_{\{\mathbf{x}_i \in A_\nu\}}\right|.$$

The first term on the right side of the inequality is bounded by $N_\nu^{-1}|\tilde{N}_\nu - N_\nu|$, which goes to zero in sample probability with $\xi$-probability 1, by Lemma A.1.

## Proof of Lemma 2

Without loss of generality we will show the result for the $l$-norm of partition created from the observed sample data $Q^{n_\nu}$ relative to $\tilde{F}_{n_\nu}$. The proof is the same for $\tilde{F}_{n_\nu}^-$.

Define $\tilde{\Delta}_i = \{\tilde{F}_{n_\nu l}(b(B^{n_\nu}(\mathbf{x}_i))) - \tilde{F}_{n_\nu l}(a(B^{n_\nu}(\mathbf{x}_i)))\}$ and $\hat{\Delta}_i = \{\hat{F}_{N_\nu l}(b(B^{n_\nu}(\mathbf{x}_i))) - \hat{F}_{N_\nu l}(a(B^{n_\nu}(\mathbf{x}_i)))\}$, so that the $l$-norms with respect to $\tilde{F}_{n_\nu}$ and $\hat{F}_{N_\nu}$ can be written as

$$\|Q^{n_\nu}\|_l^{\tilde{F}_{n_\nu}} = \tilde{N}_\nu^{-1} \sum_{i \in U_\nu} \tilde{\Delta}_i \delta_i \pi_{\nu i}^{-1}$$

and

$$\|Q^{n_\nu}\|_l^{\hat{F}_{N_\nu}} = N_\nu^{-1} \sum_{i \in U_\nu} \hat{\Delta}_i,$$

respectively.

The difference

$$\left| \|Q^{n_\nu}\|_l^{\tilde{F}_{n_\nu}} - \|Q^{n_\nu}\|_l^{\hat{F}_{N_\nu}} \right|$$

$$= \left| \tilde{N}_\nu^{-1} \sum_{i \in U_\nu} \tilde{\Delta}_i \delta_i \pi_{\nu i}^{-1} - N_\nu^{-1} \sum_{i \in U_\nu} \hat{\Delta}_i \right|$$

$$\leq \left| (\tilde{N}_\nu^{-1} - N_\nu^{-1}) \sum_{i \in U_\nu} \tilde{\Delta}_i \delta_i \pi_{\nu i}^{-1} \right|$$

$$+ \left| N_\nu^{-1} \sum_{i \in U_\nu} \{\tilde{\Delta}_i \delta_i \pi_{\nu i}^{-1} - \hat{\Delta}_i\} \right|$$

$$= |1 - (\tilde{N}_\nu/N_\nu)| \|Q^{n_\nu}\|_l^{\tilde{F}_{n_\nu}}$$

$$+ \left| N_\nu^{-1} \sum_{i \in U_\nu} \{\tilde{\Delta}_i \delta_i \pi_{\nu i}^{-1} - \hat{\Delta}_i\} \right|.$$

Now $|1 - (\tilde{N}_\nu/N_\nu)| \|Q^{n_\nu}\|_l^{\tilde{F}_{n_\nu}} \leq |1 - (\tilde{N}_\nu/N_\nu)|$ which goes to zero in probability by Lemma A.1. The quantity $|N_\nu^{-1} \sum_{i \in U_\nu} \{\tilde{\Delta}_i \delta_i \pi_{\nu i}^{-1} - \hat{\Delta}_i\}|$ is

$$\leq \left| N_\nu^{-1} \sum_{i \in U_\nu} (\tilde{\Delta}_i - \hat{\Delta}_i) \delta_{\nu i} \pi_{\nu i}^{-1} \right| + \left| N_\nu^{-1} \sum_{i \in U_\nu} (\delta_{\nu i} \pi_{\nu i}^{-1} - 1) \hat{\Delta}_i \right|,$$

where $N_\nu^{-1} \sum_{i \in U_\nu} (\delta_{\nu i} \pi_{\nu i}^{-1} - 1) \hat{\Delta}_i$ goes to zero in probability by Lemma A.2, since $\hat{\Delta}_i$ is conditionally independent of the sample given $Q^{n_\nu}$.

The quantity $|N_\nu^{-1} \sum_{i \in U_\nu} (\tilde{\Delta}_i - \hat{\Delta}_i) \delta_{\nu i} \pi_{\nu i}^{-1}|$ is

$$\leq \left( N_\nu \min_{i \in U_\nu} \pi_{\nu i} \right)^{-1} n_\nu \max_{i \in U_\nu} |\tilde{\Delta}_i - \hat{\Delta}_i| = \max_{i \in U_\nu} |\tilde{\Delta}_i - \hat{\Delta}_i| O_p(1)$$

because of Condition 2. Next we show that $\max_{i \in U_\nu}(\tilde{\Delta}_i - \hat{\Delta}_i)$ goes to zero in probability which will prove the lemma.

For a given $i \in U_\nu$ the quantity $|\tilde{\Delta}_i - \hat{\Delta}_i| = O_p(n^{-1/2})$ as a direct result of Lemma A.1. Note that $|\tilde{\Delta}_i - \hat{\Delta}_i|$ only depends on the box created from the algorithm containing $\mathbf{x}_i$. Thus we can write the quantity

$$\max_{i \in U_\nu} (\tilde{\Delta}_i - \hat{\Delta}_i) = \max_{B^{n_\nu}(\mathbf{x}_i) \in Q^{n_\nu}} \left( \tilde{\Delta}_{B^{n_\nu}(\mathbf{x}_i)} - \hat{\Delta}_{B^{n_\nu}(\mathbf{x}_i)} \right)$$

$$\leq \sum_{B^{n_\nu}(\mathbf{x}_i) \in Q^{n_\nu}} \left| \tilde{\Delta}_{B^{n_\nu}(\mathbf{x}_i)} - \hat{\Delta}_{B^{n_\nu}(\mathbf{x}_i)} \right|.$$

The second term on the right side also goes to zero in sample probability with $\xi$-probability 1. This follows from Lemma A.2 because $\mathbb{1}_{\{\mathbf{x}_i \in A_\nu\}}$ is conditionally independent of the sample given $Q^{n_\nu}$.

Since the recursive partitioning algorithm requires that there be at least $k(n_\nu)$ sample elements in each box, there are at most $n_\nu/k(n_\nu)$ boxes, so

$$\sum_{B^{n_\nu}(\mathbf{x}_i) \in Q^{n_\nu}} \left| \tilde{\Delta}_{B^{n_\nu}(\mathbf{x}_i)} - \hat{\Delta}_{B^{n_\nu}(\mathbf{x}_i)} \right| = O_p\left(n_\nu^{-1/2}\{k(n_\nu)^{-1}\}\right).$$

The definition of the function $k$ requires that $n_\nu^{1/2}\{k(n_\nu)^{-1}\}$ go to zero.

## APPENDIX B: ILLUSTRATION OF DIFFERENCES BETWEEN TWO NORMS

The following simple example illustrates the difference between the measures $\|Q^{N_\nu}\|_l^F$ and $\|Q^{N_\nu}\|_l^{F^-}$. This gives motivation for using both measures in the conditions of the main result.

*Example B.1.* Let $X_l$ be a random variable that takes one of the following three values with given probabilities:

$$X_l = \begin{cases} 1, & \text{with probability } 0.5 \\ 2, & \text{with probability } 0.25 \\ 3, & \text{with probability } 0.25. \end{cases}$$

Suppose $Q^{N_\nu}$ consists of two boxes $B_1^{N_\nu}$ and $B_2^{N_\nu}$ defined by a single split on only one $X_l$,

$$B_1^{N_\nu} = \{i \in U_\nu | x_{il} \leq 2.5\} \quad \text{and} \quad B_2^{N_\nu} = \{i \in U_\nu | x_{il} > 2.5\}.$$

Then

$$\|Q^{N_\nu}\|_l^F = \sum_{k=1,2} \left[F_l(b_l(B_k^{N_\nu})) - F_l(a_l(B_k^{N_\nu}))\right] P(B_k^{N_\nu})$$

$$= [F_l(2) - F_l(1)] P(B_1^{N_\nu}) + [F_l(3) - F_l(2)] P(B_2^{N_\nu})$$

$$= [3/4 - 1/2]\frac{3}{4} + [1 - 3/4]\frac{1}{4} = \frac{1}{4}$$

and

$$\|Q^{N_\nu}\|_l^{F^-} = [F_l^-(2) - F_l^-(1)] P(B_1^{N_\nu}) + [F_l^-(3) - F_l^-(2)] P(B_2^{N_\nu})$$

$$= [1/2 - 0]\frac{3}{4} = \frac{1}{4} + [3/4 - 1/2]\frac{1}{4} = \frac{7}{16}.$$

## REFERENCES

Benedetti, R., Espa, G., and Lafratta, G. (2008), "A Tree-Based Approach to Forming Strata in Multipurpose Business Surveys," *Survey Methodology*, 34, 195–203. [1626]

Breidt, F. J., and Opsomer, J. D. (2009), "Nonparametric and Semiparametric Estimation in Complex Surveys," in *Sample Surveys: Theory, Methods and Inference, Handbook of Statistics*, Vol. 29, eds. C. R. Rao and D. Pfeffermann, New York: North-Holland. [1626]

Breidt, J., and Opsomer, J. (2000), "Local Polynomial Regression Estimators in Survey Sampling," *The Annals of Statistics*, 28, 1026–1053. [1626]

Breidt, J., Claeskens, G., and Opsomer, J. (2005), "Model-Assisted Estimation for Complex Surveys Using Penalized Splines," *Biometrika*, 92, 831–846. [1626]

Breiman, L., Friedman, J. H., Olshen, R. A., and Stone, C. J. (1984), *Classification and Regression Trees*, New York: Chapman & Hall. [1633]

De'ath, G. (2002), "Multivariate Regression Trees: A New Technique for Modeling Species-Environment Relationships," *Ecology*, 83, 1105–1117. [1626]

De'ath, G., and Fabricius K. E. (2000), "Classification and Regression Tress: A Powerful Yet Simple Technique for Ecological Data Analysis," *Ecology*, 81, 3178–3192. [1626]

Dorfman, A., and Hall, P. (1993), "Estimators of the Finite Population Distribution Function Using Nonparametric Regression," *The Annals of Statistics*, 21, 1452–1475. [1626]

Edwards, T., Cutler, D., Zimmermann, N., Geiser, L., and Moisen, G. (2006), "Effects of Sample Survey Design on the Accuracy of Classification Tree Models in Species Distribution Models," *Ecological Modelling*, 199, 132–141. [1626]

Fuller, W. A. (1976), *Introduction to Statistical Time Series*, New York: Wiley. [1634]

Göksel, H., Judkins, D., and Mosher, W. (1992), "Nonresponse Adjustment for a Telephone Follow-up to a National in-Person Survey," *Journal of Official Statistics*, 8, 417–431. [1626]

Gordon, L., and Olshen, R. (1978), "Asymptotically Efficient Solutions to the Classification Problem," *The Annals of Statistics*, 6, 515–533. [1626-1628, 1633]

——— (1980), "Consistent Nonparametric Regression From Recursive Partitioning Schemes," *Journal of Multivariate Analysis*, 10, 611–627. [1626-1628, 1633]

Holt, D., Smith, T., and Winter, P. (1980), "Regression Analysis of Data From Complex Surveys," *Journal of the Royal Statistical Society, Ser. A*, 143, 474–487. [1626]

Horvitz, D. G., and Thompson, D. J. (1952), "A Generalization of Sampling Without Replacement From a Finite Universe," *Journal of the American Statistical Association*, 47, 663–685. [1634]

Isaki, C., and Fuller, W. A. (1982), "Survey Design Under the Regression Superpopulation Model," *Journal of the American Statistical Association*, 77, 89–96. [1626, 1628]

John, G. (1995), "Robust Decision Trees: Removing Outliers From Databases," in *Proceedings of the First International Conference on Knowledge Discovery and Data Mining*, Menlo Park, CA: AIII Press, pp. 174–179. [1629]

Lobell, D., Ortiz-Monasterio, I., Asner, G., Naylor, R., and Falcon, W. (2005), "Combining Field Surveys, Remote Sensing and Regression Trees to Understand Yield Variations in an Irrigated Wheat landscape," *Agronomy Journal*, 97, 241–249. [1626]

Mendez, G., Buskirk, T., Lohr, S., and Haag, S. (2008), "Factors Associated With Persistence in Science and Engineering Majors: An Exploratory Study Using Classification Trees and Random Forests," *Journal of Engineering Education*, 97, 57–70. [1626]

Morgan, J., and Sonquist, J. (1963), "Problems in the Analysis of Survey Data, and a Proposal," *Journal of the American Statistical Association*, 58, 415–434. [1626]

Mukhopadhyay, P. (2006), "On Some Asymptotic Results in Survey Sampling" *Pakistan Journal of Statistics*, 22, 185–193. [1634]

Nathan, G., and Holt, D. (1980), "The Effect of Survey Design on Regression Analysis," *Journal of the Royal Statistical Society, Ser. B*, 42, 377–386. [1626]

Robinson, P. M. (1982), "On the Convergence of the Horvitz–Thompson Estimator," *Australian Journal of Statistics*, 24, 234–238. [1634]

Robinson, P. M., and Särndal, C. (1983), "Asymptotic Properties of the Generalized Regression Estimator in Probability Sampling," *Sankhya, Ser. B*, 45, 240–248. [1626, 1628, 1629, 1634]

Simpson, P., Gossett, J., Parker, J., and Hall, R. (2003), "Logistic Regression Modeling—JMP Start (TM) Your Analysis With a Tree," in *Proceedings of SAS Users Group International Meeting*, Cary, NC: SAS Institute, Paper 257-28. [1626]

Stone, C. (1977), "Consistent Nonparametric Regression," *The Annals of Statistics*, 5, 595–620. [1626]

Zheng, H., and Little, R. (2004), "Penalized Spline Nonparametric Mixed Models for Inference About a Finite Population Mean From Two-Stage Samples," *Survey Methodology*, 30, 209–218. [1626]